\documentclass[conference]{IEEEtran}
\IEEEoverridecommandlockouts
\usepackage{cite}
\usepackage{amsmath,amssymb,amsfonts}
\usepackage{algorithmic}
\usepackage{graphicx}
\usepackage{textcomp}
\usepackage{xcolor}
\usepackage{tabularx}
\usepackage{caption}

\renewcommand\IEEEkeywordsname{Keywords}
\usepackage{booktabs} 
\usepackage{subcaption}
\usepackage{makecell}
\usepackage{booktabs}
\usepackage{threeparttable}
\usepackage{caption}
\def\BibTeX{{\rm B\kern-.05em{\sc i\kern-.025em b}\kern-.08em
    T\kern-.1667em\lower.7ex\hbox{E}\kern-.125emX}}

\begin{document}

\title{Exploring Wavelet Transformations for Deep Learning-based Machine Condition Diagnosis\\
\thanks{This study is supported by ASTI-GAA Project MaSense of DOST-ASTI.}
}

\author{
    \IEEEauthorblockN{Eduardo Jr Piedad}
    \IEEEauthorblockA{DOST--Advanced Science and Technology Institute\\
    Quezon City, Philippines\\
    Universitat Politècnica de Catalunya, Barcelona, Spain\\
    eduardojr.piedad@asti.dost.gov.ph}
    \and
    \IEEEauthorblockN{Christian Ainsley Del Rosario}
    \IEEEauthorblockA{DOST--Advanced Science and Technology Institute (DOST-ASTI)\\
    Quezon City, Philippines\\
    ainsley.drosario@gmail.com}
    \and
    \IEEEauthorblockN{Eduardo Prieto-Araujo}
    \IEEEauthorblockA{\textit{CITCEA, Departament d’Enginyeria Elèctrica}\\
    Universitat Politècnica de Catalunya, Barcelona, Spain\\
    eduardo.prieto-araujo@upc.edu}
    \and
    \IEEEauthorblockN{Oriol Gomis-Bellmunt}
    \IEEEauthorblockA{\textit{CITCEA, Departament d’Enginyeria Elèctrica}\\
    Universitat Politècnica de Catalunya, Barcelona, Spain\\
    oriol.gomis@upc.edu}
}

\IEEEoverridecommandlockouts
\IEEEpubid{\makebox[\columnwidth]{979-8-3503-6149-0/24/\$31.00~\copyright2024 IEEE \hfill}
\hspace{\columnsep}\makebox[\columnwidth]{ }}

\maketitle
\IEEEpubidadjcol

\begin{abstract}

Deep learning (DL) strategies have recently been utilized to diagnose motor faults by simply analyzing motor phase current signals, offering a less costly and non-intrusive alternative to vibration sensors. This research transforms these time-series current signals into time-frequency 2D representations via Wavelet Transform (WT). The dataset for motor current signals includes 3,750 data points across five categories: one representing normal conditions and four representing artificially induced faults, each under five different load conditions: 0, 25, 50, 75, and 100\%. The study employs five WT-based techniques—WT-Amor, WT-Bump, WT-Morse, WSST-Amor, and WSST-Bump. Subsequently, five DL models adopting prior Convolutional Neural Network (CNN) architecture were developed and tested using the transformed 2D plots from each method. The DL models for WT-Amor, WT-Bump, and WT-Morse showed remarkable effectiveness with peak model accuracy of 90.93, 89.20, and 93.73\%, respectively, surpassing previous 2D-image-based methods that recorded accuracy of 80.25, 74.80, and 82.80\% respectively using the identical dataset and validation protocol. Notably, the WT-Morse approach slightly exceeded the formerly highest ML technique, achieving a 93.20\% accuracy. However, the two WSST methods that utilized synchrosqueezing techniques faced difficulty accurately classifying motor faults. The performance of Wavelet-based deep learning methods offers a compelling alternative for machine condition monitoring.

\end{abstract}


\section{Introduction}
Fault detection in machines is crucial for maintaining efficiency in industrial operations. Swift and precise identification of mechanical faults is essential for minimizing production downtime and facilitating immediate remedial actions to avert further delays and disruptions. The application of artificial intelligence (AI) in this field has been marked by its capability for advanced pattern recognition and predictive analytics without the need for explicit programming. The scholarly reviews by \cite{lei2020review} and \cite{cen2022review} highlight the contributions of AI in detecting motor faults. Innovations in this area include the transformation of time-series motor current signals into 2D occurrence and recurrence plots, as introduced in recent studies of \cite{FOPCNN,nandi2019diagnosis} and \cite{9281699} which developed AI models based on convolutional neural networks (CNNs), achieving a commendable accuracy of 82.80\%. Nonetheless, conventional ML techniques of \cite{briza2024simpler} that apply a 1D frequency-transformed dataset can still demonstrate superior performance, recording an accuracy of 93.20\% using the same dataset and AI model configuration. The ongoing exploration for more effective methods indicates that 2D image-based transformations hold promising potential for enhancing fault detection accuracy.

Wavelet transform is acclaimed for its ability to provide 2D image time-frequency representations of time-series signals, which allows for a thorough investigation of complex patterns hidden within the signals. By decomposing signals at various scales, this method uncovers critical details often concealed in raw data, as noted in seminal works by \cite{daubechies1990wavelet} and \cite{mallat1989theory}. These detailed wavelet-based features enhance the diagnostic capabilities of convolutional neural networks (CNNs) for detecting motor faults. Studies of \cite{yang2018wavelet, zhang2017deep}, and more recently \cite{martinez2019wavelet, liu2020advanced} have shown how integrating wavelet-based features with CNNs leads to superior fault detection in motors, thereby underscoring the robustness of this methodology. 

This study extends the utilization of wavelet transform, exploring its variants, such as synchrosqueezed techniques, to convert motor current signal datasets into 2D time-frequency plots. These plots serve as inputs for developing CNN models for classifying motor faults. Its findings are compared with prior studies \cite{FOPCNN,9849605,nandi2019diagnosis}, employing the same motor dataset and identical CNN architectures for model development. This comparison evaluates the effectiveness of various wavelet-based transformations. The succeeding sections discuss the motor dataset and WT, the CNN model development, the results and discussion, and the conclusion and recommendation.

\section{Motor Dataset}
The single-phase electric current signals dataset was gathered from five 2-HP induction motors, comprising one normal condition, while the four others were subjected to artificial faults —- bearing axis misalignment, stator inter-turn short circuit, broken rotor bar, and outer bearing fault. These were tested under five distinct load conditions -- 0, 25, 50, 75, and 100\%, as in the methodologies of prior studies \cite{FOPCNN,9849605,nandi2019diagnosis}. Each motor was monitored over a five-second interval at a sampling rate of 10kHz, creating a total of 3,750 signal samples. Subsequently, this data was processed into 2D time-frequency plots using wavelet transform and then fed into each class's CNN model.

    \begin{figure}[tb]
            \centering
            \includegraphics[width=1\linewidth]{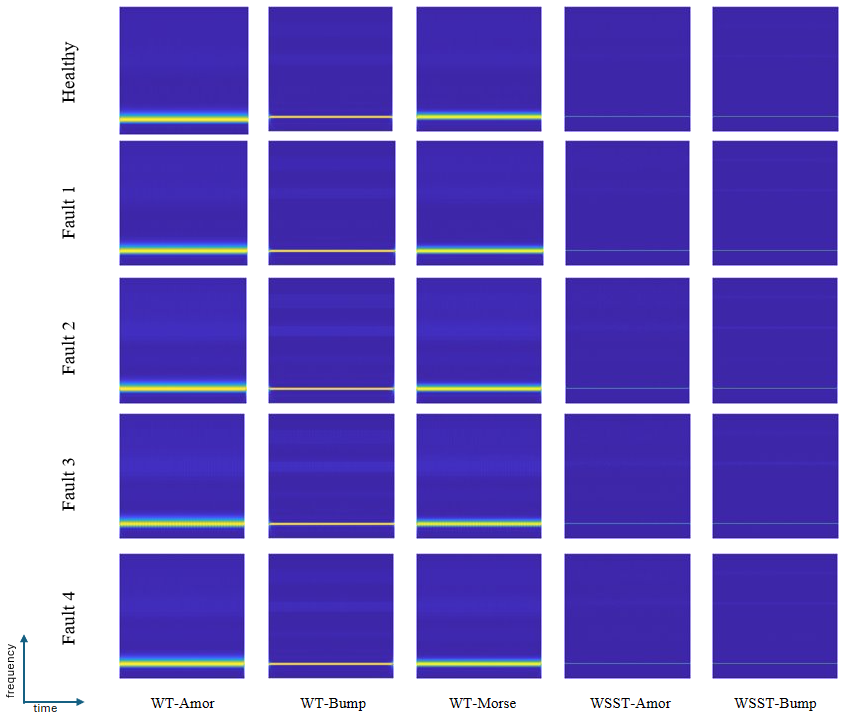}
            \caption{Sample plots showing the five Wavelet transform variants with 32x32 image resolution of the motor dataset across five condition classes under the full-load condition (load=100\%).}
            \label{wt_samples}
    \end{figure} 
    \begin{figure}[tb]
        \centering
        \includegraphics[width=1\linewidth]{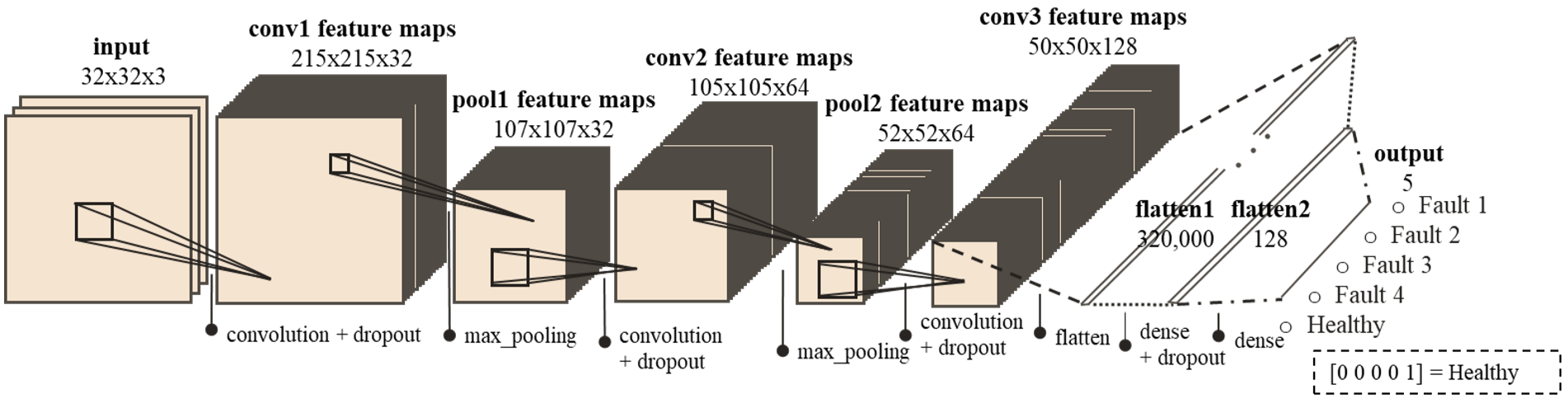}
        \caption{The Convolutional Neural Network (CNN) architecture of \cite{FOPCNN} with the following input layer corresponding to the 32x32 RGB images.}
        \label{cnn_architecture}
    \end{figure}
    \begin{figure} [tb]
        \centering
        \includegraphics[width=1\linewidth]{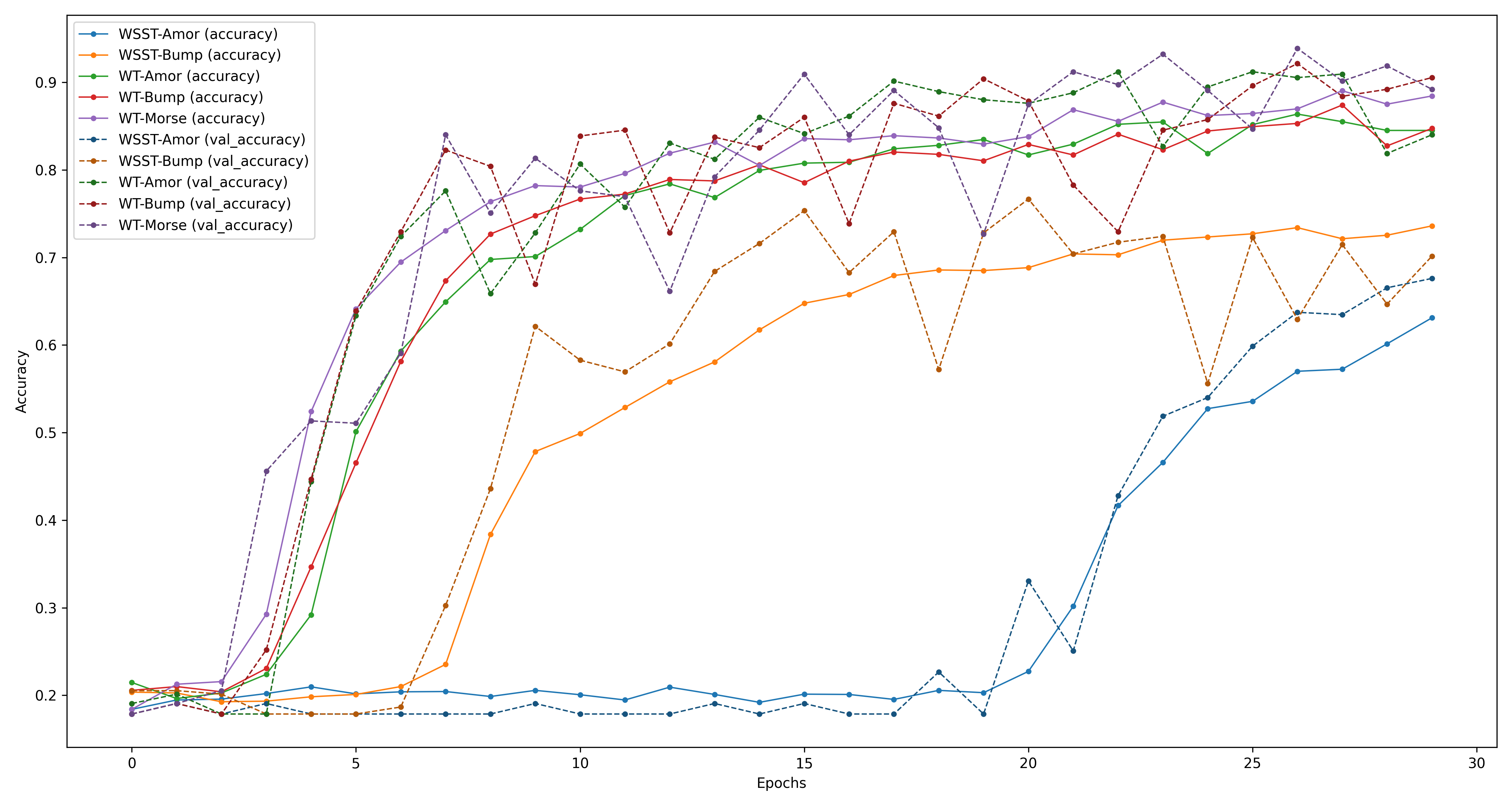}
        \caption{The training and validation classification accuracy performances of the five WT models}
        \label{training_validation_accuracy}
    \end{figure}
    
    \begin{figure}[tb]
        \centering
        \includegraphics[width=1\linewidth]{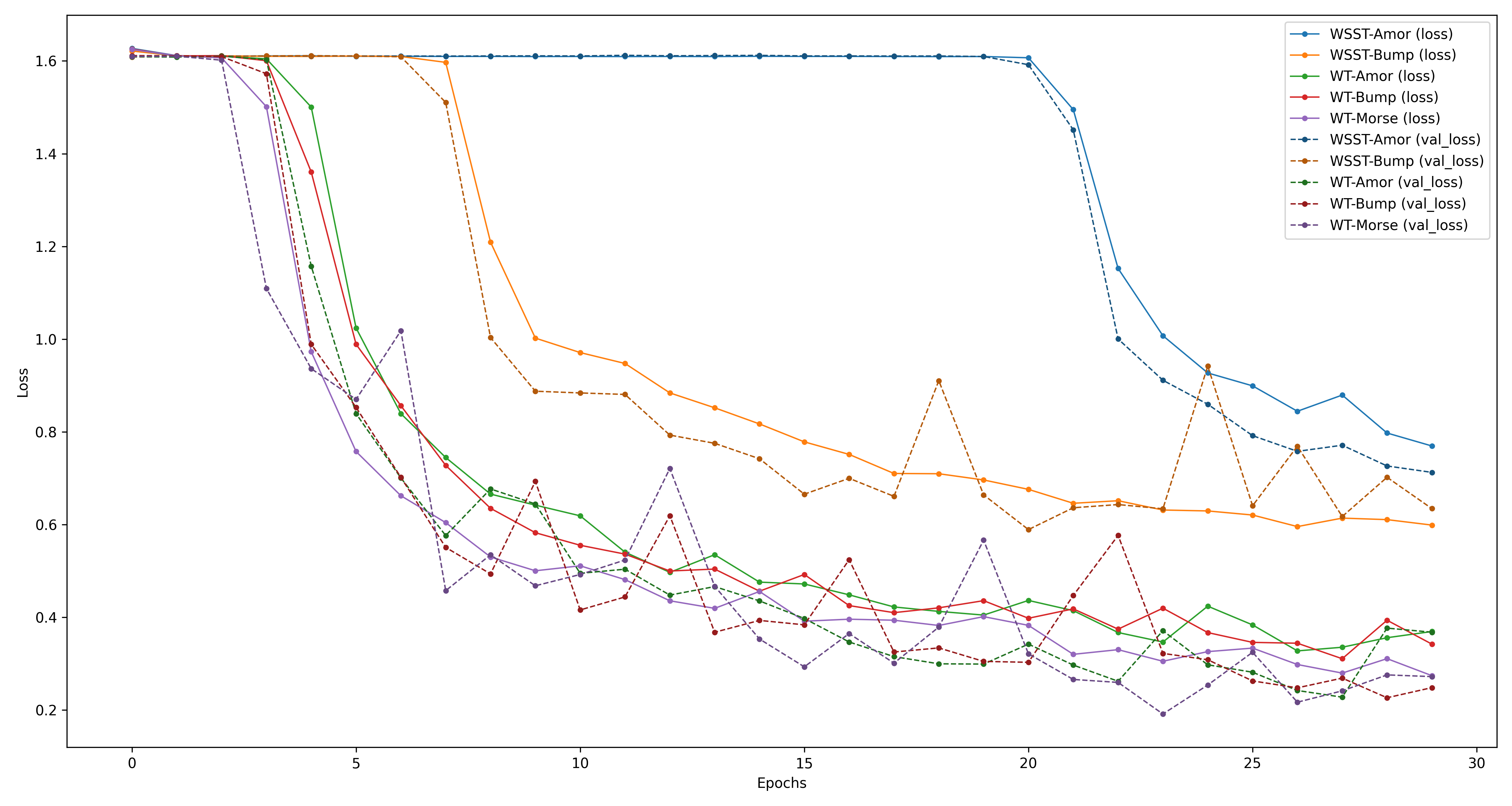}
        \caption{The training and validation loss function graphs of the five WT models.}
        \label{training_validation_loss}
    \end{figure}
   
    \begin{figure}[tb]
            \centering
            \includegraphics[width=1\linewidth]{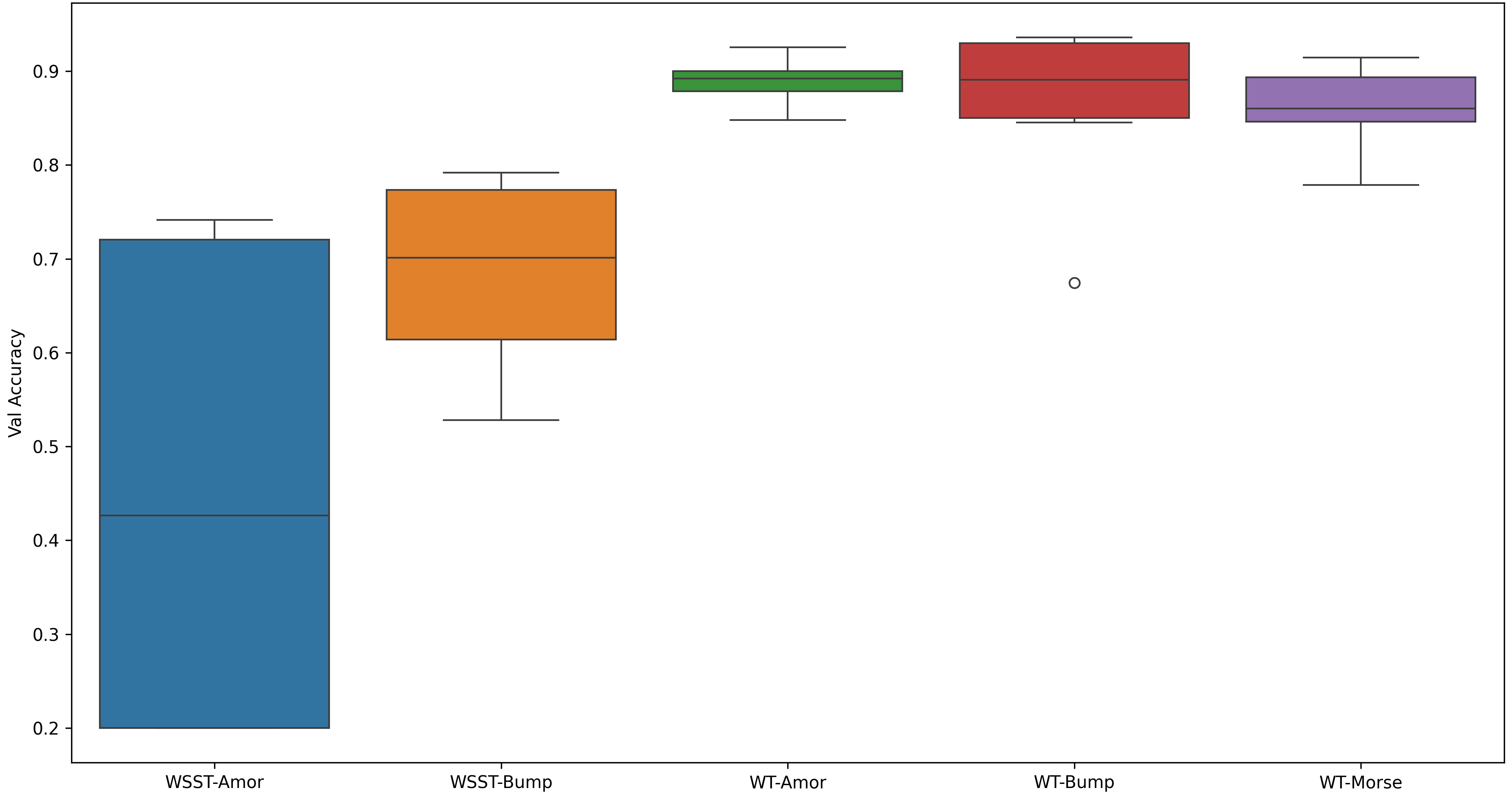}
            \caption{A box plot of the five WT models performances under the 10-fold stratified cross validation step}
            \label{cross_val}
    \end{figure}
    \begin{figure*}[tb]
            \centering
            \includegraphics[width=1\linewidth]{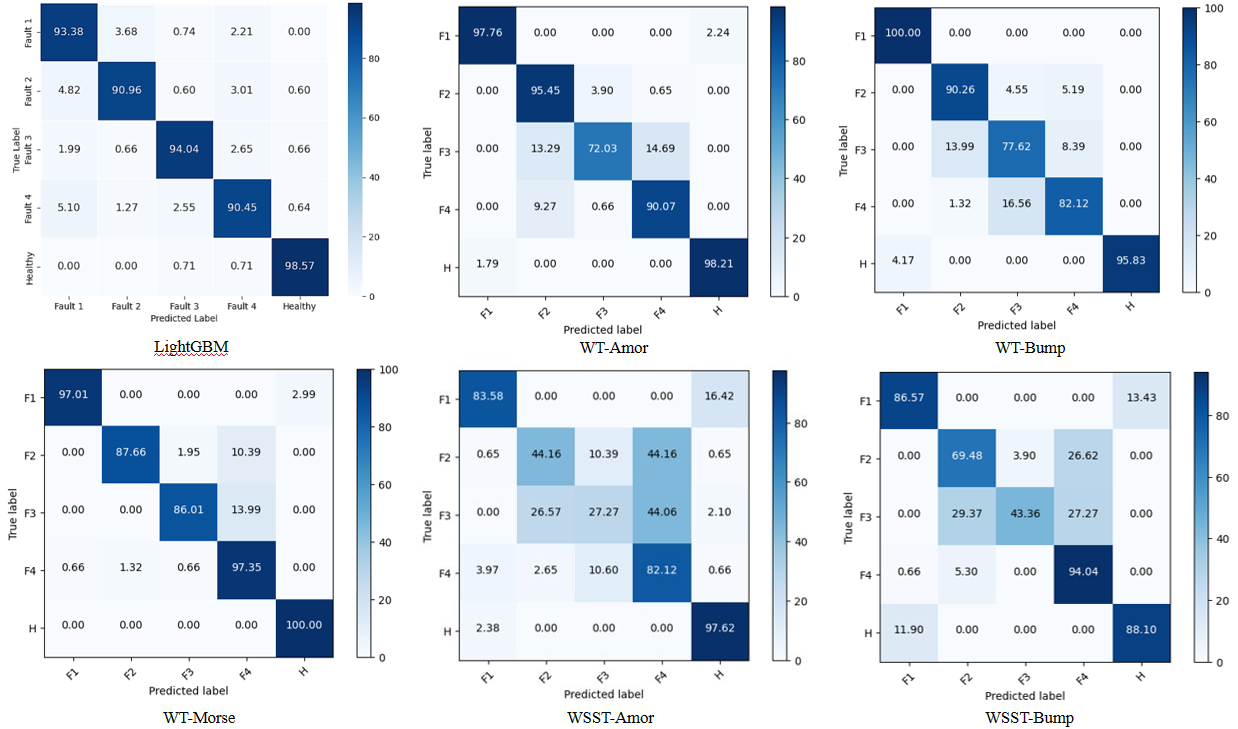}
            \caption{The confusion matrix performance of the previous best machine learning model, LightGBM, and the best model of each WT method.}
            \label{old_vs_wt}
    \end{figure*}
    
\section{Wavelet Transform and its variants}
Wavelet Transform (WT) is an essential time-frequency analysis technique for signal processing. The fundamental equation for WT is expressed as follows:
    \begin{equation}\label{wt}
        WT\{x(t)\}(a, b) = \frac{1}{\sqrt{a}} \int_{-\infty}^{\infty} x(t) \psi\left(\frac{t-b}{a}\right) dt
    \end{equation}
Here, \( \psi(t) \) represents the mother wavelet function, \( a \) is the scale parameter, and \( b \) is the translation parameter. This formulation allows the wavelet to be dilated or contracted and shifted across the time axis, enabling the extraction of signal features at different scales and positions \cite{Daubechies1992, Mallat1989}.

Variants of the Wavelet Transform, such as the Amor, Bump, and Morse wavelets, and synchrosqueezing technique, provide characteristics suited for particular signal analysis needs. The Amor wavelet is particularly effective in analyzing sinusoidal components within signals, making it a preferred choice for continuous frequency detection \cite{Farge1992}. The Bump wavelet is noted for its exceptionally smooth and rapidly decaying profile, ideal for detecting sharp transients and anomalies in signals \cite{Nason1995}. Morse wavelets are highly regarded for their tunability and asymmetry, which make them well-suited for analyzing non-stationary signals \cite{Lilly2010}. Lastly, the synchrosqueeze technique is an advanced modification of WT that refines the time-frequency representation by reallocating coefficients to more accurately reflect energy concentrations in the time-frequency domain \cite{Thakur2013}.

\section{Convolutional Neural Networks (CNNs)}
A CNN model architecture of \cite{FOPCNN} illustrated in Fig. \ref{cnn_architecture} is used in this study to allow benchmarking from its previous results. Five identical CNN models receive inputs from each WT method with three-color (Red-Green-Blue) 32x32 resolution images as shown in Fig. \ref{wt_samples}. Moreover, the methodology uses training and testing phases similar to the configurations of \cite{FOPCNN}, including the 10-fold cross-validation process.

\section{Results and Discussion} 

The two developed CNN models corresponding to the two synchrosqueezed-based WT methods were found to be difficult to converge. In contrast, the other three models converged early, as shown in their respective training and validation accuracy and loss functions in Figs. \ref{training_validation_accuracy} and \ref{training_validation_loss}, respectively. It seems that the CNN architecture works well with non-synchrosqueezed WT methods. The performances of the two synchrosqueezed-based WT methods in the 10-fold stratified cross-validation stage showed inconsistency and huge variations, as shown in Fig. \ref{cross_val}. On the other hand, WT-Amor, WT-Bump, and WT-Morse showed consistency and smaller variations after the 10-fold validation phase.

    \begin{table}[tb]
        \centering
        \begin{threeparttable}[b]
        \caption{Accuracy comparison of the best performing models of WT and prior methods on the same dataset}
        \label{table_acc}
        \begin{tabular}{llc}
        \toprule
        Method & Code & Accuracy (\%) \\
        \midrule
        \textbf{Wavelet Morse} & \textbf{WT-Morse} & \textbf{93.73} \\
        LightGBM & LGBM & 93.20 \\
        \textbf{Wavelet Amor} & \textbf{WT-Amor} & \textbf{90.93} 
        \\
        K-Nearest Neighbors & KNN & 89.90 \\
        \textbf{Wavelet Bump} & \textbf{WT-Bump} & \textbf{89.20} 
        \\
        Random Forest & RF & 88.00 \\
        AdaBoost & ADA & 84.50 \\
        Multi Layer Perceptron & MLP & 84.10 \\
        Pretrained CNN on FOP \cite{nandi2019diagnosis}  & PTCNN-FOP & 82.80 \\
        Logistic Regression & LR & 80.50 \\
        CNN on FOP \cite{FOPCNN} & CNN-FOP & 80.25 \\
        CNN on RP & CNN-RP & 74.80 \\        
        \textbf{synchrosqueezed-WT Bump} & \textbf{WSST-Bump} & \textbf{76.66}
        \\
        Support Vector Machines & SVM & 74.50 \\
        \textbf{synchrosqueezed-WT Amor} & \textbf{WSST-Amor} & \textbf{67.60}
        \\
        Decision Tree & DT & 66.00 \\
        Extra Tree & XT & 64.30 \\
        Gaussian Naive Bayes & GNB & 47.50 \\
        \bottomrule
        \end{tabular}
        \end{threeparttable}
    \end{table}

Table \ref{table_acc} summarizes the best performing models of WT methods and the prior studies of \cite{FOPCNN, 9849605,nandi2019diagnosis, briza2024simpler}. Only the WT-Morse method slightly outperformed the previous best model, while the WT-Amor and WT-Bump methods performed comparatively well with the leading models. However, the synchrosqueezed models showed unsatisfactory performance. This may be due to possible excessive manipulation or "synchrosqueezing" of the Wavelet plots compared to the normal WT variants as contrasted in Fig. \ref{wt_samples}, thus losing the underlying patterns of each class.

Moreover, Fig. \ref{old_vs_wt} shows the confusion matrices of LightGBM of \cite{briza2024simpler} and the five WT methods. The LightGBM performed well with at least 90\% accuracy across all five motor classes. At the same time, WT-Morse excelled better in classifying bearing axis misalignment and outer bearing faults but compromised the other faults - stator inter-turn short circuit and broken rotor bar, with only 87.66 and 86.01\%, respectively. WT-Amor and WT-Bump performed well only in classifying motors with bearing axis misalignment faults and normal conditions. All three performed better than the previous 2D-image-based techniques. Lastly, the synchrosqueezed-based WT performed poorly across the motor classes.

\section{Conclusion} 
Applying Wavelet Transform (WT) to convert time-series motor current signals into time-frequency 2D plots has effectively uncovered underlying features essential for predicting motor faults. The study tested five different WT-based transformation methods using convolutional neural networks (CNNs), a deep learning architecture. The WT-Amor, WT-Bump, and WT-Morse methods demonstrated significant effectiveness, showing superior performance compared to previous 2D-image-based techniques. The WT-Morse model surpassed prior methodologies and marginally outperformed the previous best model. However, the two synchrosqueezed-WT models faced substantial inconsistencies and difficulties, suggesting that the "synchrosqueezing" process might have overly manipulated the WT plots, potentially obscuring critical patterns necessary for the deep learning of CNNs. These findings show the potential of wavelet-based 2D transformations for intelligent motor fault diagnosis. 

\bibliographystyle{IEEEtran}
\bibliography{refs.bib}

\begin{thebibliography}{10}
\providecommand{\url}[1]{#1}
\csname url@samestyle\endcsname
\providecommand{\newblock}{\relax}
\providecommand{\bibinfo}[2]{#2}
\providecommand{\BIBentrySTDinterwordspacing}{\spaceskip=0pt\relax}
\providecommand{\BIBentryALTinterwordstretchfactor}{4}
\providecommand{\BIBentryALTinterwordspacing}{\spaceskip=\fontdimen2\font plus
\BIBentryALTinterwordstretchfactor\fontdimen3\font minus \fontdimen4\font\relax}
\providecommand{\BIBforeignlanguage}[2]{{%
\expandafter\ifx\csname l@#1\endcsname\relax
\typeout{** WARNING: IEEEtran.bst: No hyphenation pattern has been}%
\typeout{** loaded for the language `#1'. Using the pattern for}%
\typeout{** the default language instead.}%
\else
\language=\csname l@#1\endcsname
\fi
#2}}
\providecommand{\BIBdecl}{\relax}
\BIBdecl

\bibitem{lei2020review}
Y.~Lei, B.~Yang, X.~Jiang, F.~Jia, N.~Li, and A.~K. Nandi, ``Applications of machine learning to machine fault diagnosis: A review and roadmap,'' \emph{Mechanical Systems and Signal Processing}, vol. 138, p. 106587, 2020.

\bibitem{cen2022review}
J.~Cen, Z.~Yang, X.~Liu, J.~Xiong, and H.~Chen, ``A review of data-driven machinery fault diagnosis using machine learning algorithms,'' \emph{Journal of Vibration Engineering \& Technologies}, vol.~10, no.~7, pp. 2481--2507, 2022.

\bibitem{FOPCNN}
E.~Piedad~Jr, Y.-T. Chen, H.-C. Chang, C.-C. Kuo \emph{et~al.}, ``Frequency occurrence plot-based convolutional neural network for motor fault diagnosis,'' \emph{Electronics}, vol.~9, no.~10, p. 1711, 2020.

\bibitem{nandi2019diagnosis}
A.~Nandi, S.~Biswas, K.~Samanta, S.~S. Roy, and S.~Chatterjee, ``Diagnosis of induction motor faults using frequency occurrence image plots—a deep learning approach,'' in \emph{2019 International Conference on Electrical, Electronics and Computer Engineering (UPCON)}.\hskip 1em plus 0.5em minus 0.4em\relax IEEE, 2019, pp. 1--4.

\bibitem{9281699}
A.~L. Moraes and R.~A.~S. Fernandes, ``Recurrence plots: A novel feature engineering technique to analyze power quality disturbances,'' in \emph{2020 IEEE Power \& Energy Society General Meeting}, 2020, pp. 1--5.

\bibitem{briza2024simpler}
A.~C. Briza, E.~Piedad, and E.~C. Peramo, ``Simpler machine learning methods outperform deep learning in motor fault detection,'' in \emph{2024 International Conference on Artificial Intelligence in Information and Communication (ICAIIC)}.\hskip 1em plus 0.5em minus 0.4em\relax IEEE, 2024, pp. 675--680.

\bibitem{daubechies1990wavelet}
I.~Daubechies, \emph{The Wavelet Transform, Time-Frequency Localization and Signal Analysis}.\hskip 1em plus 0.5em minus 0.4em\relax IEEE, 1990.

\bibitem{mallat1989theory}
S.~Mallat, ``A theory for multiresolution signal decomposition: The wavelet representation,'' \emph{IEEE Transactions on Pattern Analysis and Machine Intelligence}, vol.~11, no.~7, pp. 674--693, 1989.

\bibitem{yang2018wavelet}
B.~Yang and M.~Liang, ``Application of wavelet transform in machine fault diagnosis,'' \emph{Journal of Sound and Vibration}, 2018.

\bibitem{zhang2017deep}
W.~Zhang and Y.~Yang, ``Deep learning for fault diagnosis using goodness-of-fit tests based on wavelet packet decomposition,'' \emph{Journal of Computational and Applied Mathematics}, 2017.

\bibitem{martinez2019wavelet}
J.~Martinez and P.~Rodriguez, ``Wavelet-based feature extraction for improved fault diagnosis in rotating machinery,'' \emph{IEEE Transactions on Industrial Electronics}, 2019.

\bibitem{liu2020advanced}
H.~Liu and Z.~Wang, ``Advanced wavelet deep learning models for fault diagnosis in gear systems,'' \emph{Mechanical Systems and Signal Processing}, 2020.

\bibitem{9849605}
Z.~Xu, D.~Yu, and Y.~Hu, ``Motor fault diagnosis method based on deep learning,'' in \emph{2022 11th International Conference of Information and Communication Technology (ICTech))}, 2022, pp. 236--239.

\bibitem{Daubechies1992}
I.~Daubechies, ``Ten lectures on wavelets,'' \emph{CBMS-NSF Regional Conference Series in Applied Mathematics}, 1992.

\bibitem{Mallat1989}
S.~Mallat, ``A theory for multiresolution signal decomposition: The wavelet representation,'' \emph{IEEE Transactions on Pattern Analysis and Machine Intelligence}, vol.~11, no.~7, pp. 674--693, 1989.

\bibitem{Farge1992}
M.~Farge, ``Wavelet transforms and their applications to turbulence,'' \emph{Annual Review of Fluid Mechanics}, vol.~24, pp. 395--458, 1992.

\bibitem{Nason1995}
G.~P. Nason, ``Wavelet thresholding in nonparametric regression,'' \emph{Biometrika}, vol.~81, no.~3, pp. 425--455, 1995.

\bibitem{Lilly2010}
J.~M. Lilly and S.~C. Olhede, ``On the analytic wavelet transform,'' \emph{IEEE Transactions on Information Theory}, vol.~56, no.~8, pp. 4135--4156, 2010.

\bibitem{Thakur2013}
G.~Thakur, E.~Brevdo, N.~S. Fuckar, and H.-T. Wu, ``Synchrosqueezing based recovery of instantaneous frequency from nonuniform samples,'' \emph{SIAM Journal on Mathematical Analysis}, vol.~43, no.~5, pp. 2078--2095, 2013.

\end{thebibliography}

\end{document}